\documentstyle[preprint,prl,aps,epsf]{revtex}
\begin{document}
\draft
\title{Conductivity sum rule, implication for in-plane dynamics and
$c$-axis response}
\author{Wonkee Kim and J. P. Carbotte}
\address{Department of Physics and Astronomy, McMaster University,
Hamilton, Ontario, Canada L8S~4M1}
\maketitle
\begin{abstract}
Recently observed $c$-axis optical sum rule violations indicate
non-Fermi liquid in-plane behavior.
For coherent $c$-axis coupling, the observed flat, nearly frequency
independent $c$-axis conductivity $\sigma_{1}(\omega)$ implies
a large in-plane scattering rate $\Gamma$ around
$(0,\pi)$ and therefore any pseudogap that might form at low 
frequency in the normal state will be smeared. On the
other hand incoherent $c$-axis coupling places no restriction on the
value of $\Gamma$ and gives a more consistent picture of the observed
sum rule violation which, we find in some cases, can be less than half.
\end{abstract}

$C$-axis electrodynamics is important in distinguishing high-$T_{c}$ cuprates
from conventional superconductors and for
understanding mechanism in the cuprates\cite{anderson}.
Since the cuprates have layered structures, interlayer coupling
between two adjacent CuO$_{2}$ planes plays an essential role, and its
exact nature impacts on
the recent experimental observation\cite{basov} of a significant violation of
the conventional 
optical sum rule of Ferrel, Glove and Tinkham\cite{fgt}.
The normalized missing spectral weight (NMSW)
is different from one and of order of a half
in several high-$T_{c}$ cuprates. 
NMSW is the difference between the area under the real part of the
conductivity 
$\sigma^{n(s)}(\omega)
\left[=\sigma^{n(s)}_{1}(\omega)+i\sigma^{n(s)}_{2}(\omega)\right]$
in the normal minus 
superconducting state, devided by the superfluid density 
which is obtained 
from $\sigma^{s}_{2}(\omega)$. 
We have pointed out in Ref.~\cite{kim}
that in order to explain NMSW of less than one,
it is necessary to consider non-Fermi liquid models for
the in-plane dynamics,
regardless of the nature of the interlayer coupling.
A possible such model introduces a pseudogap in the normal state
above $T_{c}$ as is observed\cite{timusk}.
Another is the "mode" coupling model determined from consideration
of ARPES data by Norman {\it et al}\cite{norman}. This model 
has been used to describe kinetic, as opposed to
potential energy driven superconductivity.

In this paper, we show that coherent $c$-axis coupling,
even with a pseudogap, cannot easily
explain recent experimental findings,
but that incoherent coupling describes them including
the observed low frequency behavior of the effective NMSW. 

The interlayer coupling is represented
by a Hamiltonian $H_{c}$\cite{radtke,hirschfeld,hirsh}:
\begin{equation}
H_{c}=\sum_{ij,\;\sigma}
\left[t_{ij}c^{+}_{i1\sigma}c_{i2\sigma}+H.c.\right]\;,
\end{equation}
where the hopping matrix $t_{ij}$ describes weak interlayer tunneling and
$c^{+}_{i1\uparrow}$ creates an electron
with spin $\uparrow $ at the site $i$ in the plane $1$.
We classify interlayer couplings in two classes 
because the results are remarkably
different depending on their nature:
i) One is coherent coupling
$(t_{ij}=t_{\perp})$, which originates from
an overlap of electronic wave functions between the two planes.
For an in-plane Fermi liquid, it was shown
in Ref. \cite{kim} that
the superfluid density $(\rho_{s})$ is equal to the missing spectral
weight $(N_{n}(\omega_{c})-N_{s}(\omega_{c})=
8\int^{\omega_{c}}_{0^{+}}d\omega'\Bigl[\sigma^{n}_{1c}(\omega')
-\sigma^{s}_{1c}(\omega')\Bigr]$, where $\omega_{c}$ is a cutoff frequency
of the order of a bandwidth).
Thus coherent coupling 
(unless the density
of states has a strong variation with energy) can explain
the experimental results on optimally doped YBCO
and over-doped Tl$2201$\cite{katz}.
ii) The other is incoherent coupling,
for which $t_{ij}=V_{i}\delta_{ij}$, where
$V_{i}$ is an impurity potential.
In this case, we showed in Ref.\cite{kim} that
NMSW $\ge 1.58$.
The characteristic difference
between coherent and incoherent coupling is whether or not
electron momentum is conserved in the interlayer transfer.

In the presence of an external vector potential $A_{z}$,
$H_{c}$ is modified to $H_{c}(A_{z})$ by 
the phase factor $\exp(\pm ieA_{z})$. 
It is sufficient to expand $H_{c}(A_{z})$ to second order in $A_{z}$.
The current $j_{c}=-\delta H_{c}(A_{z})/\delta A_{z}
=j_{p}+j_{d}$, where
$j_{p}=-ied\sum_{i\sigma}t_{\perp}\bigl[c^{+}_{i1\sigma}c_{i2\sigma}-
c^{+}_{i2\sigma}c_{i1\sigma}\bigr]$ and $j_{d}=e^{2}d^{2}H_{c}A_{z}$
with $d$ the interlayer spacing.
In linear response theory, $\langle j_{c}\rangle=
\bigl[-\Pi+e^{2}d^{2}\langle H_{c}\rangle\bigr]A_{z}$,
where $\Pi$ is
the current-current correlation function associated with $j_{p}$
and $\langle H_{c}\rangle$ is the perturbation of $j_{d}$
due to $H_{c}$.

From the $c$-axis conductivity sum rule
\cite{hirsh,klein}
the superfluid density $\rho_{s}$ can be written as
\begin{equation}
\rho_{s}=8\int^{\omega_{c}}_{0^{+}}{}d\omega\Bigl[\sigma^{n}_{1c}(\omega)
-\sigma^{s}_{1c}(\omega)\Bigr]-4\pi e^{2}d^{2}
\Bigl[\langle H_{c}\rangle^{s}-\langle H_{c}\rangle^{n}\Bigr],
\label{rho}
\end{equation}
where
$\omega_{c}$ is the cutoff frequency for 
the interband transitions that $H_{c}$
does not describe, and
we use units such that $\hbar=c=k_{B}=1$ and set the volume
of the system to be unity.

The penetration depth $\lambda_{c}$ can be calculated in two ways.
Based on the Kramers-Kronig relation for the conductivity,
we obtain $\lambda_{c}$ as
$1/4\pi\lambda^{2}_{c}=\lim_{\omega\rightarrow 0}
[\omega\mbox{I}\mbox{m}\sigma_{c}(0,\omega)]$.
Alternatively, using Eq.(\ref{rho})
we can also calculate $\lambda_{c}(=1/\sqrt{\rho_{s}})$.
Combining these two equations for $\lambda_{c}$, we obtain the formula:
\begin{equation}
{(N_{n}-N_{s})\over\rho_{s}}={1\over2}+
{1\over2}{\sum_{\omega}\sum_{{\bf k},{\bf p}}
|t_{{\bf k}-{\bf p}}|^{2}[G({\bf k},\omega)G({\bf p},\omega)-
G_{0}({\bf k},\omega)G_{0}({\bf p},\omega)]\over
\sum_{\omega}\sum_{{\bf k},{\bf p}}
|t_{{\bf k}-{\bf p}}|^{2}F({\bf k},\omega)F^{+}({\bf p},\omega)}\;,
\label{spect}
\end{equation}
where $G({\bf k},\omega)$ and $F({\bf k},\omega)$ are superconducting
Green functions and $G_{0}({\bf k},\omega)$ is in the normal state.
For coherent $c$-axis coupling 
$(|t_{{\bf k}-{\bf p}}|^{2}=t^{2}_{\perp}\delta_{{\bf k}-{\bf p}})$
electron momentum parallel to the plane is conserved and
$t^{2}_{\perp}$ may depend on the in-plane momentum.
For incoherent coupling, an impurity configuration average is implied
over a potential $V_{i}$ 
$(|t_{{\bf k}-{\bf p}}|^{2}=|V_{{\bf k}-{\bf p}}|^{2})$, and
electron momentum is not conserved, and ${\bf k}$ and ${\bf p}$ remain
unconstraint.

{\it Coherent coupling}. 
As the simplest case, one can consider a normal state
spectral function
$A_{0}({\bf k},\omega)=\Gamma/
\left[(\omega-\xi_{\bf k})^{2}+\Gamma^{2}\right]$ based on Fermi-liquid
theory. The in-plane scattering rate $\Gamma$ is expected to depend 
strongly on direction of the momentum ${\bf k}$ in the two dimensional
Brillouin zone. Cold spot exists along $(\pi,\pi)$ and 
hot spot along $(0,\pi)$\cite{norman}. The coherent $c$-axis matrix
element $t_{\perp}$, however, is itself dependent on direction and 
$t^{2}_{\perp}$ varies as $\cos^{4}(2\phi)$, where $\phi$ is the angle 
defining the in-plane direction of ${\bf k}$. This factor makes the 
$c$-axis conductivity sensitive mainly to the hot spot, although it
remains Drude-like. Here we can ignore both $\phi$ dependences
and interprete $\Gamma$ as the scattering rate coming from $(0,\pi)$
region (hot spot). However, except for YBCO at optimum doping, 
$\sigma_{1}(\omega)$ vs $\omega$ is found 
to be almost flat over an energy range
of a few $100$ meV. 
This implies that $\Gamma$ and $\Gamma_{s}$ need to be several
$100$ meV. Non-Fermi liquid models with large $\Gamma$
have appeared in the literature based on ARPES data in 
Bi$2212$. Norman {\it et al}.\cite{norman}
have derived a simple phenomenological model - called the mode model
in which the normal state is described by a large frequency independent
$\Gamma$ and the superconducting state by 
$\Gamma(\omega)=\Gamma~\theta(\omega-\Delta(0)-\omega_{mode})$,
{\it i.e.} a Fermi
liquid is recovered in this state as $\Gamma(\omega)=0$
for $\omega < \Delta(0)+\omega_{mode}$, where $\Delta(0)$ is the gap
and $\omega_{mode}$ is the frequency of some collective
mode. 
In addition, a pseudogap $\tilde{\Delta}$
can be introduced in the normal state as in the 
preformed pair model\cite{emery} with no long range phase
coherence above $T_{c}$.
We take the pseudogap to have the same $d$-wave
symmetry as does the superconducting gap in agreement with experiment
\cite{timusk} but it may not have the same amplitude. 
A possible phenomenological
form of spectral function 
$\tilde{A} ({\bf k},\omega)$ is
$\tilde{A}_{+}({\bf k},\omega)+\tilde{A}_{-}({\bf k},\omega)$, where
$\tilde{A}_{\pm}({\bf k},\omega)=
\Gamma~(1\pm\xi_{\bf k}/\tilde{E}_{\bf k})/
\left[(\omega\mp\tilde{E}_{\bf k})^{2}+\Gamma^{2}\right]$.
Here $\tilde{E}_{\bf k}=\sqrt{\xi^{2}_{\bf k}+
\tilde{\Delta}^{2}_{\bf k}}$ and the factors 
$(1\pm\xi_{\bf k}/\tilde{E}_{\bf k})$ do not imply the phase
coherence of BCS theory.

If the frequency cut-off $\omega_{c}$ 
is much larger than any energy scale in our consideration, 
then
under the assumption of a cylindrical
Fermi surface, it can be shown that as $T\rightarrow0$,
$-T\sum_{\omega,{\bf k}}G({\bf k},\omega)^{2}
=1/2+{\bf K}\left(i{\Delta(0)/\Gamma_{s}}\right)/\pi$
and
$T\sum_{\omega,{\bf k}}F({\bf k},\omega)^{2}=
1/2-{\bf K}\left(i{\Delta(0)/\Gamma_{s}}\right)/\pi$
for superconducting
state with an in-plane scattering rate $\Gamma_{s}$,
which is assumed, for simplicity, to be frequency independent, 
and 
$-T\sum_{\omega,{\bf k}}G_{0}({\bf k},\omega)^{2}
=1/2+{\bf K}\left(i{\tilde{\Delta}/\Gamma}\right)/\pi$
for normal pseudogap
state. Here ${\bf K}$ is the complete elliptic integral of
the first kind and
we can set the density of states $N(0)$ at the Fermi level equal to be unity
because it does not appear in NMSW.
The in-plane angle dependence of $t^{2}_{\perp}$ 
such as $\cos^{4}(2\phi)$ 
does not change the results
as long as $\omega_{c}\gg \Delta(0)$ and $\tilde{\Delta}$. 
Therefore, NMSW becomes
\begin{equation}
{(N_{n}-N_{s})\over\rho_{s}}={1\over2}+
{{\bf K}\left(i{\tilde{\Delta}\over\Gamma}\right)
-{\bf K}\left(i{\Delta(0)\over\Gamma_{s}}\right)\over{\pi-
2\,{\bf K}\left(i{\Delta(0)\over\Gamma_{s}}\right)}}\;.
\label{cnmsw}
\end{equation}
$\Gamma_{s}$ may be different
from $\Gamma$ and much smaller as ARPES data imply\cite{timusk}. 
Such data are more consistent with the existence of quasiparticles
in the superconducting state than in the normal state.

It is worthwhile illustrating the
implications of Eq.~(\ref{cnmsw}) in some detail.
i) Suppose the in-plane scattering rate $\Gamma_{s}$
is negligible
compared with $\Delta(0)$, then Eq.~(\ref{cnmsw}) reduces
to $(N_{n}-N_{s})/\rho_{s}=
1/2+{\bf K}\left(i{\tilde{\Delta}/\Gamma}\right)/\pi$,
which depends only on the normal state and 
NMSW $\ge 1/2$. In this case, the conventional sum rule is recovered
when $\tilde{\Delta}/\Gamma\ll 1$ because the spectral function
$\tilde{A} ({\bf k},\omega)$ becomes insensitive to the pseudogap. 
ii) If $\Delta(0)/\Gamma_{s}\gg 1$ and $\tilde{\Delta}/\Gamma\gg 1$, then 
NMSW $\simeq 1/2$. To get $1/2$, it is not necessary
that $\Delta(0)\simeq\tilde{\Delta}$ and $\Gamma_{s}\simeq\Gamma$
simultaneously;
in other words, a mismatch between $\tilde{\Delta}$ and $\Delta(0)$
does not matter as long as $\Gamma$ and $\Gamma_{s}$ are 
both negligible. 
But a small value of $\Gamma$ gives 
a frequency dependence of $\sigma^{n}_{1c}(\omega)$, which disagrees
with the observation that it is flat for $\omega$ up to
a few $100$ meV. Thus, the large $\Gamma$ and $\Gamma_{s}$ limit must be
applied, and consequently any pseudogap region at low 
$\omega\lesssim\tilde{\Delta}$ will be filled in and the model cannot 
describe the observation made in underdoped YBCO. 
A cancelation between $\sum_{\omega,{\bf k}}G({\bf k},\omega)^{2}$
and $\sum_{\omega,{\bf k}}G_{0}({\bf k},\omega)^{2}$
in Eq.~(\ref{spect}) can still arise
for large $\Gamma$ if there is
a match between $\Delta(0)$ and $\tilde{\Delta}$ and between
$\Gamma$ and $\Gamma_{s}$ as can be seen from Fig.1, where we plot
$-T\sum_{\omega,{\bf k}}G_{0}({\bf k},\omega)^{2}$ (or
$-T\sum_{\omega,{\bf k}}G({\bf k},\omega)^{2}$) for three values of the
pseudogap (or gap) as a function of $\Gamma$ (or $\Gamma_{s}$). 
We also show $T\sum_{\omega,{\bf k}}F({\bf k},\omega)^{2}$.
Clearly the large $\Gamma$ region  
can also give a value of the sum rule
bigger or smaller than $1/2$. Should
the pseudogap be larger than the superconducting gap 
(point ${\bf b}$ in Fig.~1)
and $\Gamma_{s}$ not too much smaller than $\Gamma$ (point ${\bf a}$),
the sum rule will be less than $1/2$ while if $\Gamma_{s}$ is
much less than $\Gamma$
such as for point ${\bf d}$, it will be larger than $1/2$ but less than $1$.
Other combinations could also be used. To be more explicit,
suppose that $\tilde{\Delta}/\Gamma=\Delta(0)/\Gamma_{s}+\gamma$
with $\Delta(0)/\Gamma_{s}\gg\gamma$, 
then ${\bf K}(i\tilde{\Delta}/\Gamma)/\pi
-{\bf K}(i\Delta(0)/\Gamma_{s})/\pi\simeq -(\gamma/4)\Delta(0)/\Gamma_{s}$.
Since $1/2-{\bf K}(i\Delta(0)/\Gamma_{s})/\pi
\simeq(1/8)(\Delta(0)/\Gamma_{s})^{2}$ for $\Delta(0)/\Gamma_{s}\ll 1$,
NMSW$\simeq1/2-\gamma\Gamma_{s}/\Delta(0)$ for 
$\gamma\ll\Delta(0)/\Gamma_{s}\ll 1$.
Based on ARPES data, we may assume that $\tilde{\Delta}\simeq\Delta(0)$ and
$\Gamma\gg\Gamma_{s}\rightarrow0$. In this case NMSW$\simeq 
1/2+({\bf a}-{\bf d})/(2{\bf c})$,
which is obviously greater than $1/2$, but small $\Gamma_{s}$ will 
not give the
observed flat response $\sigma^{s}_{1c}(\omega)$ in the superconducting state.

So far we have assumed that both
superconducting and normal state are at $T\rightarrow0$.
Because it is not possible to access the normal state at 
low $T$,
Basov {\it et al.} have used $T=T_{c}$ instead.
In the inset of Fig.~1,
we plot $-T\sum_{\omega,{\bf k}}G^{2}_{0}$ 
with $\tilde{\Delta}=4T_{c}$ and $\Gamma=T_{c}$ as a function of $T$.
(We emphasize that
$-T\sum_{\omega,{\bf k}}G^{2}_{0}$
is nothing but NMSW if $\Gamma_{s}\ll\Delta(0)$.)
It is clear that
including the $T$ dependence in the NMWS
does not qualitatively change the physics of
the conductivity sum rule.

{\it Incoherent coupling}.
For incoherent $c$-axis coupling,
we need a specific model for the impurity scattering potential 
$|V_{{\bf k}-{\bf p}}|^{2}$ and need to 
average over impurity configuration.
Part of the disorder scattering can be due to a mismatch of
overlap matrix elements $t_{\perp}$ between planes.
We use a simple model for the scattering potential
$|V_{{\bf k}-{\bf p}}|^{2}=|V_{0}|^{2}+|V_{1}|^{2}\cos(2\phi_{k})
\cos(2\phi_{p})$\cite{hirschfeld} and assume
$|V_{0}|^{2}=|V_{1}|^{2}$ for simplicity. The NMSW is now 
\begin{equation}
{(N_{n}-N_{s})\over\rho_{s}}={1\over2}+
{1\over2}
{\sum_{\omega}\left[\tilde{\kappa}'^{2}{\bf K}^{2}(\tilde{\kappa})-
\kappa'^{2}{\bf K}^{2}(\kappa)\right]\over
{\sum_{\omega}\left[(\kappa'^{2}/\kappa){\bf K}(\kappa)
-{\bf E}(\kappa)/\kappa\right]^{2}}}\;,
\label{inmsw}
\end{equation}
where
$\tilde{\kappa}=\tilde{\Delta}/
\sqrt{\tilde{\Delta}^{2}+(\omega+\Gamma\mbox{sgn}\omega)^{2}}$,
$\tilde{\kappa}'=\sqrt{1-\tilde{\kappa}^{2}}$, 
$\kappa=\Delta(T)/\sqrt{\Delta(T)^{2}
+(\omega+\Gamma_{s}\mbox{sgn}\omega)^{2}}$,
$\kappa'=\sqrt{1-\kappa^{2}}$,
and ${\bf E}$ is 
the complete elliptic integral of the second kind.
As one can easily see, $\tilde{\kappa}'^{2}{\bf K}^{2}(\tilde{\kappa})$
becomes $(\pi/2)^{2}$ if $\tilde{\Delta}=0$ regardless of
$\Gamma$. It can also be inferred from Eq.~(\ref{inmsw}) that
i) as $\tilde{\Delta}$ increases or $\Gamma$ decreases,
NMSW becomes smaller because $\tilde{\kappa}$ is closer to
$1$ for a given $\omega$, and ii) as $\Gamma_{s}$ increases
NMSW becomes smaller because $\kappa$ is closer to $0$.
In the incoherent coupling case, the normal state response
$\sigma^{n}_{1c}\propto |V_{0}|^{2}$ is independent of $\omega$
and there is no restriction
on $\Gamma$\cite{kim,hirschfeld}.
In the inset of Fig.~2, we plot NMSW as a function of $\Gamma_{s}$
with fixed values of $\alpha=0$ and $1$, where
$\alpha\equiv\tilde{\Delta}/\Delta(0)$, and
$\Gamma=\Delta(0)$. As expected, NMSW with 
$\alpha=1$ decreases with increasing
$\Gamma_{s}$. Note that NMSW with $\alpha=0$ is almost unchanged with
increasing $\Gamma_{s}$ because the numerator and the denominator 
of the second term in Eq.~(\ref{inmsw}) are
both decreasing in similar manner.
In the case $\tilde{\Delta}=\Delta(0)(1+\delta)$ 
$\left[\delta\ll 1\right]$ 
with $\Gamma=\Gamma_{s}=0$, Eq.~(\ref{inmsw}) reduces to
NMSW$\simeq1/2-1.083\,\delta$, which is less than $1/2$.
For $\Gamma=\Gamma_{s}\ne0$, the coefficient of $\delta$ will be changed.
In the general case we need to
evaluate the sums in Eq.~\ref{inmsw}
for given $\tilde{\Delta}$, $\Delta(0)$, $\Gamma$ and $\Gamma_{s}$.
In the main frame of Fig.~2, 
we plot NMSW as a function of $\Gamma$ with a few values
of $\alpha$ assuming $\Gamma_{s}=0$ since we have seen that
$\Gamma_{s}\ne0$ decreases NMSW.
As one sees, NMSW is greater than 1 and independent of $\Gamma$ if $\alpha=0$.
Note that NMSW with $\alpha=0$ is an asymptote of NMSW with nonzero $\alpha$
because the effect of $\tilde{\Delta}$ on NMSW vanishes
when $\Gamma\gg\tilde{\Delta}$. 
As we expect, NMSW decreases with increasing
$\alpha$ and decreasing $\Gamma$.
If $\alpha\ge1.46$ and 
$\Gamma\ll\Delta(0)$, then NMSW$< 0$. Generally speaking,
in the incoherent coupling model,
we can explain NMSW$\approx 1/2$ with
$\tilde{\Delta}\approx\Delta(0)$, $\Gamma > \Delta(0)$ and
$\Gamma_{s} < \Delta(0)$. 

Additional insight into the sum rule can be obtained
from consideration of $\sigma^{n(s)}_{1c}(\omega)$. In the top panel of
Fig.~3, we compare
$\sigma^{n}_{1c}(\omega)$ (solid curve), with 
pseudogap
$\tilde{\Delta}=\Delta(0)$, with $\sigma^{s}_{1c}(\omega)$
(dashed curve). In both cases, $\Gamma$ is taken to be zero and therefore
both curves go to zero at $\omega=0$. At small $\omega\lesssim\Delta(0)$
the solid and dashed curves are almost the same. 
However, they start to differ at a higher $\omega$, with 
$\sigma^{s}_{1c}(\omega)$ falling below $\sigma^{n}_{1c}(\omega)$.
We already know that for these parameter the sum rule is exactly $1/2$.
But in this example, the effective NMSW defined as 
${\cal N}(\omega)=\left[N_{n}(\omega)-N_{s}(\omega)\right]/\rho_{s}$
is nealy zero up to $\omega\gtrsim\Delta(0)$ as shown in the dashed curve
in the bottom panel of Fig.~3. 
This does not agree with the findings of Basov {\it et al.},
where ${\cal N}(\omega)$ increases almost linearly out of
zero and rapidly move toward its saturated value. It is clear that,
even for the incoherent $c$-axis coupling case, the argument
presented in Ref.\cite{millis} that the preformed pair model
agrees with a sum rule of $1/2$ due to a simple cancelation between the
normal and superconducting state
is deficient when its detailed approach toward
its saturated value is considered, {\it i.e.} when we compare 
the low $\omega$ behavior of ${\cal N}(\omega)$ with experiment.

A more reasonable model is obtained
for finite normal state $\Gamma$. Returning
to the solid curve of the top panel 
and including a finite $\Gamma$ in the 
pseudogap state fills in the region below $2{\tilde \Delta}$ by
transfering spectral weight out of the region above this. In the middle
panel we show results (solid curve) for a pseudogap ${\tilde \Delta}=
1.4\Delta(0)$ and $\Gamma=0.5\Delta(0)$ for illustration only.
Other choices could also have been made showing that our explanations
are robust. Comparison of the solid curve (normal state)
with the superconducting (dashed) curve $(\Gamma_{s}=0)$ shows a very
different low $\omega$ behavior. Now ${\cal N}(\omega)$ will grow rapidly 
as $\omega$ increases from $\omega=0$. This behavior is shown as the solid 
curve of the bottom panel. ${\cal N}(\omega)$ grows linearly out of
$\omega=0$ and nearly reaches its saturated value$(\simeq0.52)$ by
$\omega\approx 6\sim7\Delta(0)$ in much better agreement with experiment.
One can also conceive that ${\tilde \Delta}$ and $\Gamma$ 
are larger than considered above and that
the normal state may not exhibit much of a pseudogap because of 
a large $\Gamma$, but that the sum rule is still 
nevertheless less than 1. 
This could explain optimally doped Tl$2201$ with NMSW$\approx 0.6$ where,
however, no in-plane pseudogap has been reported
although in the data of Basov {\it et al.}, $\sigma^{n}_{1c}(\omega)$
is droping with
decreasing $\omega$ at small $\omega$ consistent with a smeared
pseudogap.

{\it Conclusion}. 
We conclude that incoherent $c$-axis coupling can
more easily explain the observed violations of the 
optical sum rule 
than can coherent coupling, but that non-Fermi liquid in-plane behavior 
is required. 
It is possible that the normalized missing spectral weight
is less than $1/2$. 
A sharp increase in 
accumulated normalized spectral weight as a function of cut-off frequency
rising to a value close to its saturated value within a few times the gap
as is observed can also be understood within our phenomenological model
but not within a pure
preformed pair model\cite{millis}.

Work supported by the Natural Sciences and Engineering
Research Council of Canada (NSERC).
W. K. acknowledges N. Whelan and P. J. Hirschfeld for discussions and
correspondence, respectively.
J. P. C thanks D. Basov for enlightening
discussions.

\begin{figure}
\caption{The quantity $-T\sum_{\omega,{\bf k}}G_{0}^{2}$ (or
$-T\sum_{\omega,{\bf k}}G^{2}$) 
as a function
of $\Gamma/T_{c}$ (or $\Gamma_{s}/T_{c}$) for different values
of ${\tilde\Delta}$ and $\Delta(0)$: ${\tilde\Delta}=T_{c}$ (dashed curve),
${\tilde\Delta}=\Delta(0)=2.5T_{c}$ (solid curve) and ${\tilde\Delta}=4T_{c}$
(dotted curve). The solid curve is $T\sum_{\omega,{\bf k}}F^{2}$.
The inset shows the $T$ dependence 
of $-T\sum_{\omega,{\bf k}}G_{0}^{2}$ with ${\tilde\Delta}=4T_{c}$ and
$\Gamma=T_{c}$.}

\caption{NMSW as a function of  $\Gamma/\Delta(0)$ for different values
of $\alpha\equiv{\tilde\Delta}/\Delta(0)$ with $\Gamma_{s}=0$ and 
$T\ll\Delta(0)$. The inset gives NMSW as a function of $\Gamma_{s}/\Delta(0)$
for two specific cases: $\alpha=0$ with any $\Gamma$ (dashed curve)
and $\alpha=1$ with $\Gamma=\Delta(0)$ (solid curve).}

\caption{Top panel shows $\sigma^{n}_{1c}(\omega)/\sigma_{cn}$ 
with ${\tilde\Delta}=
\Delta(0)$ and $\Gamma=0$ (solid curve),
where $\sigma_{cn}=4\pi n_{i}\left(edN(0)V_{0}\right)^{2}$ with
an impurity density $n_{i}$. 
The dashed curve 
$(\sigma^{s}_{1c}(\omega)/\sigma_{cn}$ is for
the superconducting state with $\Delta(0)$ and $\Gamma_{s}=0$.
Middle panel gives $\sigma^{n}_{1c}(\omega)/\sigma_{cn}$ but now with
${\tilde\Delta}=1.4\Delta(0)$ and $\Gamma=0.5\Delta(0)$. Note 
the transfer of spectral weight from higher to lower frequency as compared
to case when $\Gamma=0$ (solid curve in the top panel). Bottom panel
shows the effective NMSW ${\cal N}(\omega)$. The dashed curve is for
the preform pair model with $\Gamma=\Gamma_{s}=0$ and 
${\tilde\Delta}=\Delta(0)$. Its saturated value of $1/2$ is not
reached at $\omega=6\Delta(0)$ where it is still only about $0.3$. 
The solid curve
includes smearing into the normal state
with ${\tilde\Delta}=1.4\Delta(0)$ and $\Gamma=0.5\Delta(0)$.
Its saturated value is about $0.52$ which is nearly reached by
$\omega=6\Delta(0)$.}
\end{figure}

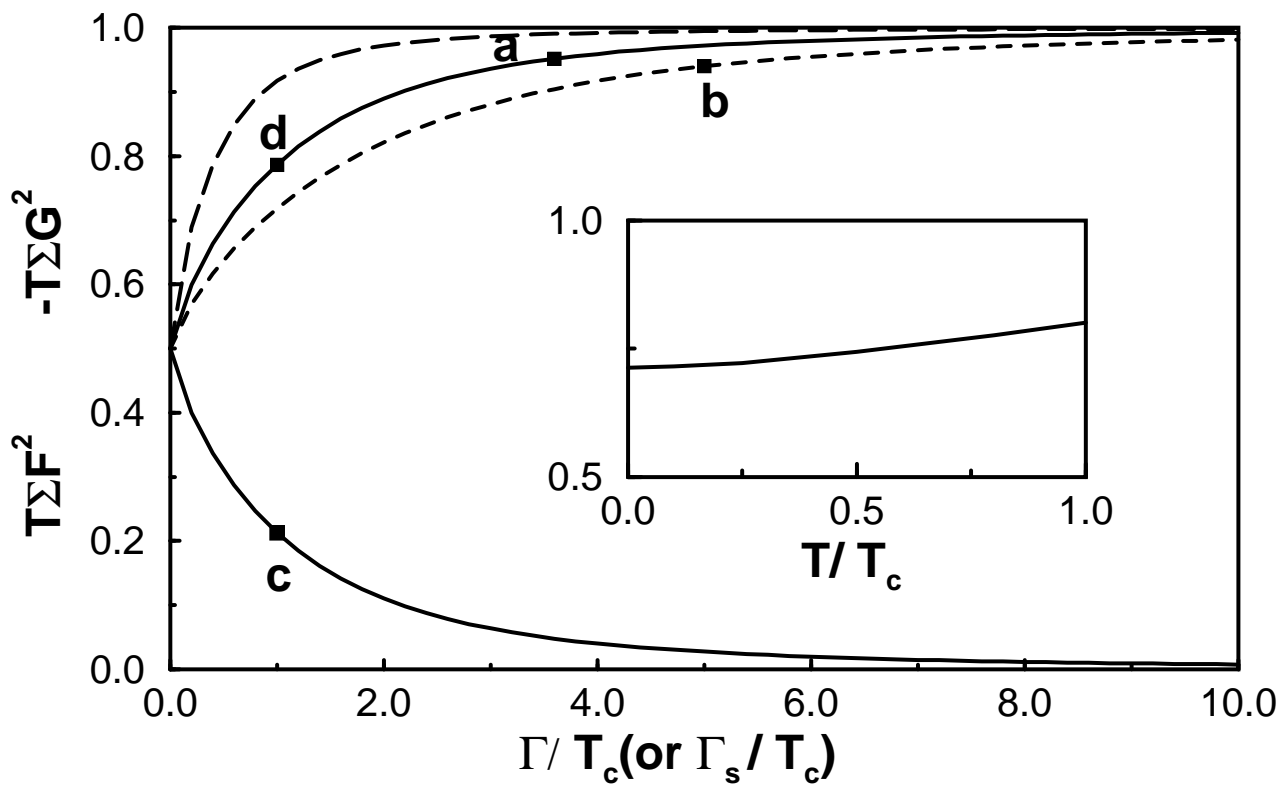
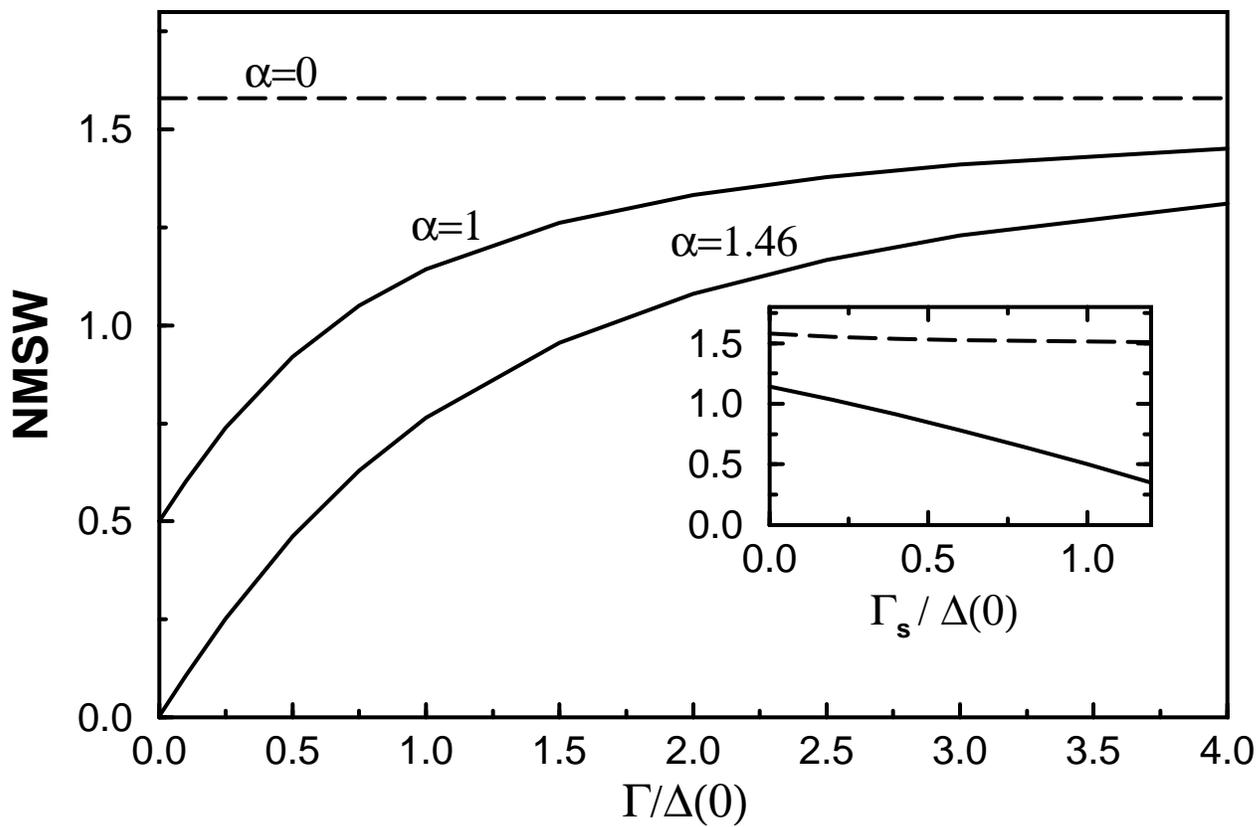
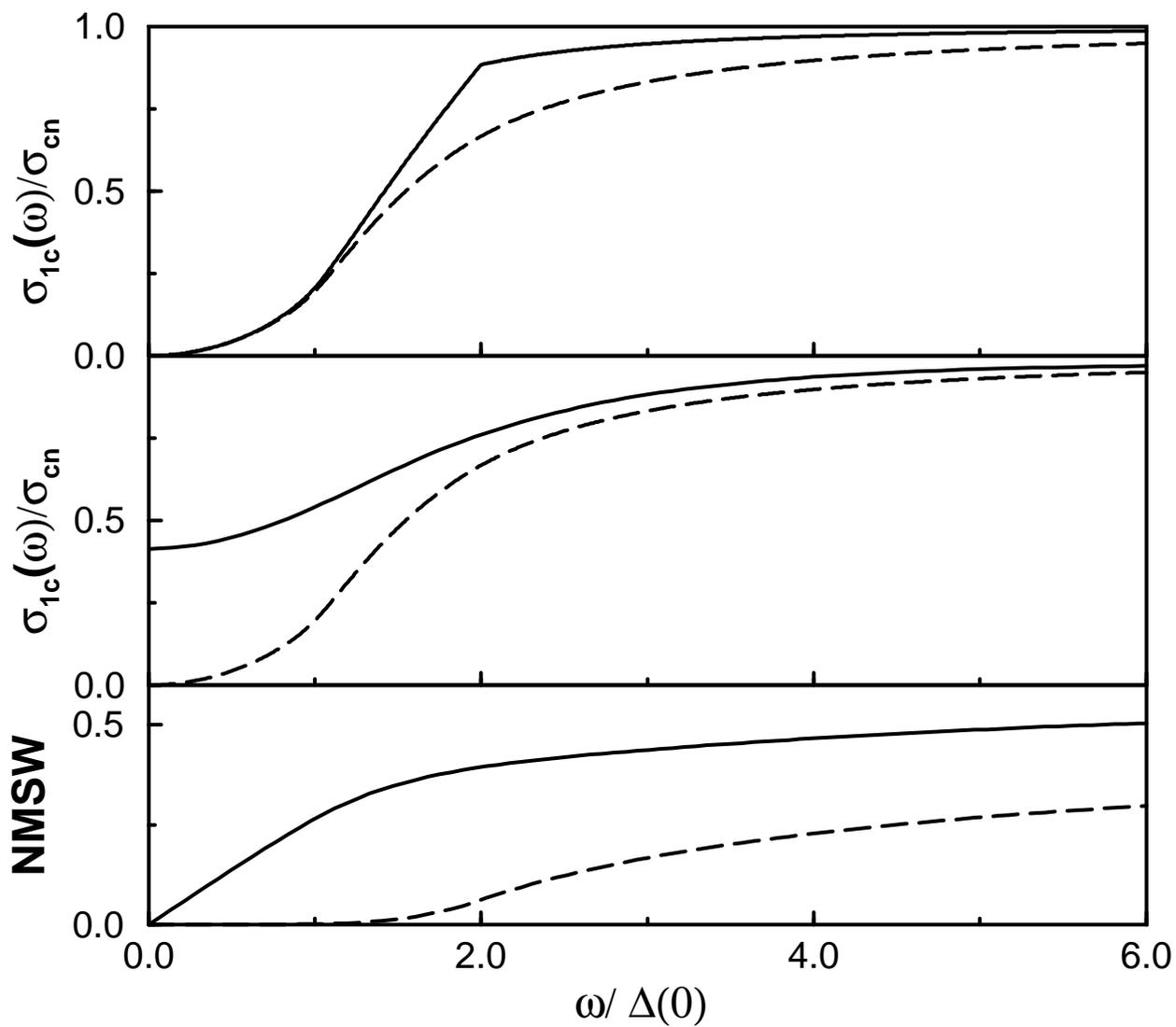
\end{document}